**RESEARCH**                                                                              **Open Access**

# Analysing gamification elements in educational environments using an existing Gamification taxonomy

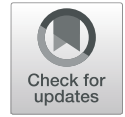

Armando M. Toda[1,2*], Ana C. T. Klock[3], Wilk Oliveira[1], Paula T. Palomino[1], Luiz Rodrigues[1], Lei Shi[2], Ig Bittencourt[5], Isabela Gasparini[4], Seiji Isotani[1] and Alexandra I. Cristea[2]

* Correspondence:
[1]Universidade de Sao Paulo - Av. Trab. Sao Carlense, 400, zip code, Sao Carlos, SP 13566-590, Brazil
[2]Durham University - Lower Mountjoy South Road Durham, zip code, Durham DH1 3LE, UK
Full list of author information is available at the end of the article

## Abstract

Gamification has been widely employed in the educational domain over the past eight years when the term became a trend. However, the literature states that gamification still lacks formal definitions to support the design and analysis of gamified strategies. This paper analysed the game elements employed in gamified learning environments through a previously proposed and evaluated taxonomy while detailing and expanding this taxonomy. In the current paper, we describe our taxonomy in-depth as well as expand it. Our new structured results demonstrate an extension of the proposed taxonomy which results from this process, is divided into five dimensions, related to the learner and the learning environment. Our main contribution is the detailed taxonomy that can be used to design and evaluate gamification design in learning environments.

## Introduction

Gamification has been extensively used in educational environments and instructional practices (Dichev and Dicheva 2017) to enhance students' engagement and motivation through the employment of game design elements outside of a fully-fledged game (Barata et al. 2015; Deterding et al. 2011; Kapp 2012; Nand et al. 2019). While recognizing the available game elements and choosing which of them must be employed in gamified environments are not trivial tasks, some gamification frameworks are aiming to help designers with that. However, many of these frameworks have no common understanding of the set of game elements that can be used by gamified systems and the knowledge on how to apply them. (Dichev and Dicheva 2017; Klock et al. 2018b; Mora et al. 2015; Toda et al. 2018b). Besides, there are no naming conventions and the process to support which elements belong to gamification are other issues found in gamification literature in general, as they use different synonyms for the same game element, e.g., badges and trophies (Koivisto and Hamari 2019; Pedreira et al. 2015; Seaborn and Fels 2014).

All these hinder the adoption of gamification by teachers and instructors, since recent studies demonstrated that these specialists have interest in using gamification but does not have time or resources to make sense of differences and similarities in deciding which game elements to use, as well as which game elements are more appropriate in educational context (Martí-Parreño et al. 2016; Sánchez-Mena and Martí-Parreño 2016; Toda et al. 2018a). Aiming to solve this problem, an initial taxonomy of game elements for gamified educational environments was proposed and evaluated [omitted for blind review]. We defined a poll of 21 game elements, alongside their synonyms, and validated them through two surveys with experts in the field of gamification in education. However, the initial taxonomy did not present how these elements could be grouped and organised in a way that could guide researchers, designers and instructors to use them more efficiently. Here, as an extension of a previous study, we propose to answer a more practical research question *"How can we use the proposed taxonomy to analyse and evaluate gamified educational environments?"*. By answering this research question, our contributions include:

- improving the existing taxonomy, by providing details on the selection, description, and use of these elements to evaluate and analyse existing systems;
- proposing recommendations on how to hierarchically organise these elements semantically, to be used by designers, teachers, and other education stakeholders.

## Related works

Gamification frameworks are not a novelty nowadays, and recent literature reviews have mapped more than 50 frameworks focused on how to design gamification in a specific or broad context (Azouz and Lefdaoui 2018; Mora et al. 2017). However, only a few of them were focused on education and learning contexts (less than 10). Following the nomenclature issue previously described, these frameworks proposed different concepts with similar descriptions: while "a title attributed to the player that he can use to compare with others" is called *Social Status* by Marczewski (2015), it is a *Classification* in Dignan (2011). In this section, we present some existing taxonomies based on their adoption and the context of the framework/taxonomy.

Concerned with general contexts, which are frameworks that were created for general purposes, we have the taxonomy proposed in six steps to gamification (6D) (Werbach and Hunter 2012) presenting a hierarchy of game elements using Dynamics, Mechanics and Components, based on the MDA framework (Hunicke et al. 2004). In this classification, the top of the hierarchy is composed of the Dynamics, which are the abstractions related to the task that is being gamified. These Dynamics used to create the motivation to perform the task and are manifested via Mechanics. The Mechanics are the processes used to drive the users' actions and are presented through the Components. Finally, these components are extrinsic rewards and feedback features like points, badges, etc. The taxonomy presented in the 6D framework, however, does not provide the user with clear strategies on how to combine these elements properly. Also, being a general framework, it lacks instances on educational environments, validated empirically.

Next, we have the GAME framework (Marczewski 2015) which provides an extensive periodic table of gamification elements ($n = 52$). Their taxonomy is divided by player profiles ($n = 8$), where these elements may work better based on the users' player profile. In this framework, we can already observe some similarities with the concepts proposed in Werbach. In the GAME framework, the Progress/Feedback is treated as a component that may engage general contexts, while in the 6D Progress is treated as a dynamic and Feedback as a Mechanic.

As for the frameworks used in the educational context, we opted to choose recent ones that were instanced.[1] The framework proposed by Klock et al. (2016) consists of 7 steps to aid the design of adaptive gamification in e-learning environments. In this framework, the authors use a set of 14 gamification elements based on 6D and apply it to develop an adaptive e-learning system. Another recent taxonomy is presented in the work of Toda et al. (2018a) where the authors propose a framework focused on teachers and instructors. The game elements in this work are divided into Feedback and Property. The Feedback elements are the ones that can be used as feedback, and the Property is characteristics and objectives for the educational task. The authors define a poll of 19 gamification elements and provide some strategies (based on existing literature) on their use. However, both taxonomies were not validated.

Finally, none of the taxonomies that were presented explicit ways on how to analyse those elements in learning environments nor how to analyse these elements. An overview can be seen in Table 1.

## Methods and tools

As explained in [**omitted for blind review**], the game elements were collected, analysed and defined by the authors, then evaluated by gamification experts.[2] The collection was based on a literature review made by the authors, where they analysed the nomenclature of other gamification frameworks and analysed the concepts that were presented. Based on semantic analysis, we defined a set of 21 gamification elements that could be used in educational systems. After the initial definition, we designed an evaluation focusing on five variables:

- **Comprehensibility**: the standardised concept for the group of game elements, the "name".
- **Description**: the concept definition.
- **Relevance**: the relevance of that element in the overall taxonomy.
- **Examples**: the examples tied to the definition and concept.
- **Coverage**: the representation of the overall taxonomy. If this set of 21 elements represent and cover well the game elements needed for educational applications.

In this paper, we focus on expanding the descriptions of the gamification elements that were presented, and choose some existing gamified educational environments, based on their popularity and presence in research papers, to analyse these elements have been applied and interpret why, since this can be used to support designers to select the most appropriate game elements in their environments. The use of the taxonomy to support the process of analysis and evaluation was supervised by five gamification experts, that would analyse the systems and match with the elements in our taxonomy.

---

[1]By instanced, we mean that they were applied and evaluated in a real educational context
[2]Most of the experts were also teachers and researchers

Table 1 Related works comparison

| Taxonomy | Field | Focus | Number of elements | Present Instances | Validation |
|---|---|---|---|---|---|
| Werbach and Hunter (2013) | General | Design | 30 | No | No |
| Marczewski (2015) | General | Design | 52 | No | No |
| Klock et al. (2016) | Education | Design | 14 | Yes | No |
| Toda et al. (2018a) | Education | Design | 19 | Yes | No |
| Actual Taxonomy | Education | Design, analysis and evaluation | 21 | Yes | Yes |

Following, we focused on expanding the concepts, by giving examples of how these elements are represented in the literature, as well as advantages and disadvantages in employing them. Finally, we proposed a new hierarchical classification for these elements, that can support designers and developers to choose which elements to use in the make of gamified strategies.[3] This classification was designed starting by identifying five dimensions, each one associated with an aspect of the environment. To design these dimensions, the concepts were analysed on a semantic level and discussed amongst at least five researchers. The complete process can be seen in Fig. 1.

## Results

This section describes the definitions of the taxonomy, some synonyms, and examples of how each element can be applied in an educational environment and some advantages and disadvantages in its use. We also propose an initial definition of extrinsic (when an element is presented in a way that the user can perceive it clearly and objectively) and intrinsic elements (which is an element presented in a subtle way that the user may not notice when interacting with the environment). An overall of the new taxonomy can be seen in Fig. 2.

### Description of the five dimensions

Extending the initial taxonomy, we propose a classification using five dimensions to group the previously defined gamification elements. Each element was analysed by at least five experts to group each in the appropriate dimension, e.g.: *When analysing the Point element, the experts noticed that it was an extrinsic feedback element, that is given when the learner executes a certain action within the environment. Since it is given to the learner as a form of feedback, it would be appropriate to classify it as part of the Performance/Measurement dimension.* We describe the dimensions as follows.

### Performance / measurement

These are elements related to the environment response, which can be used to provide feedback to the learner. In this dimension we have Point, Progression, Level, Stats and Acknowledgement. Lack of this dimension means that the student may feel disoriented as their actions does not have any kind of feedback.

---
[3]A Gamified strategy is a task, with a goal, that contains game-like elements (A. M. Toda, do Carmo, et al., 2018).

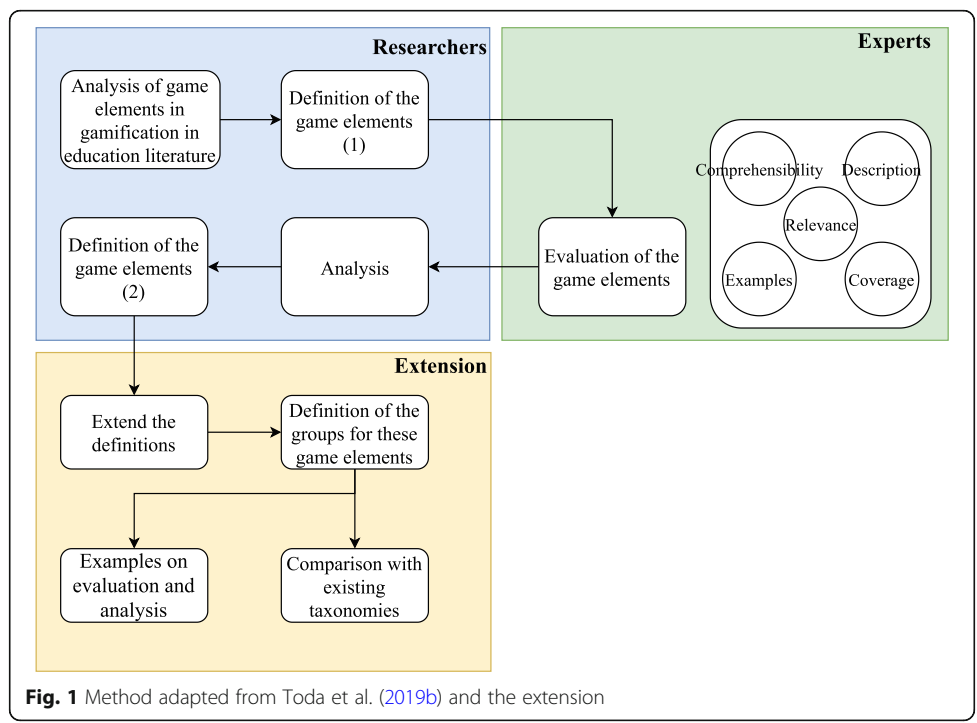

**Fig. 1** Method adapted from Toda et al. (2019b) and the extension

- **Acknowledgement:** also known as badges, medals, trophies and achievements. It is a kind of extrinsic feedback that praises the players' specific set of actions, e.g. completing a certain number of problems may lead them to earn a "Solver" badge; finishing a task in a predefined time limit may earn them a "Flash" trophy; making a certain number of interactions with other students may give them a "Socialiser" achievement; making a certain number of contributions may earn them a "Contributor" badge. Acknowledgement is one of the most used elements in gamified applications (Klock et al. 2018a; Koivisto and Hamari 2019; Toda et al. 2018b).
- **Level:** also known as skill level, character level etc. This is related to an extrinsic hierarchical layer that provides the user new advantages as they advance in the environment, e.g. the students gain a level every time they complete a certain number of tasks, when they advance their level, they have access to more challenging tasks.
- **Progression**: also known as progress bars, steps, maps. Provides an extrinsic guidance to the users of their advance in the environment, allowing these users to locate themselves.
- **Point**: also known as scores, experience points, skill points, etc. It is a simple way to provide extrinsic feedback to the users' actions. Point is the most basic concept found in almost all gamified applications (Dichev and Dicheva 2017).
- **Stats:** also known as information, Head Up Display (HUD) and data. It is related to the visual information provided by the environment to the learner (extrinsic), e.g. how many tasks they completed or overall stats on the environment. In virtual environments this can also be dashboards.

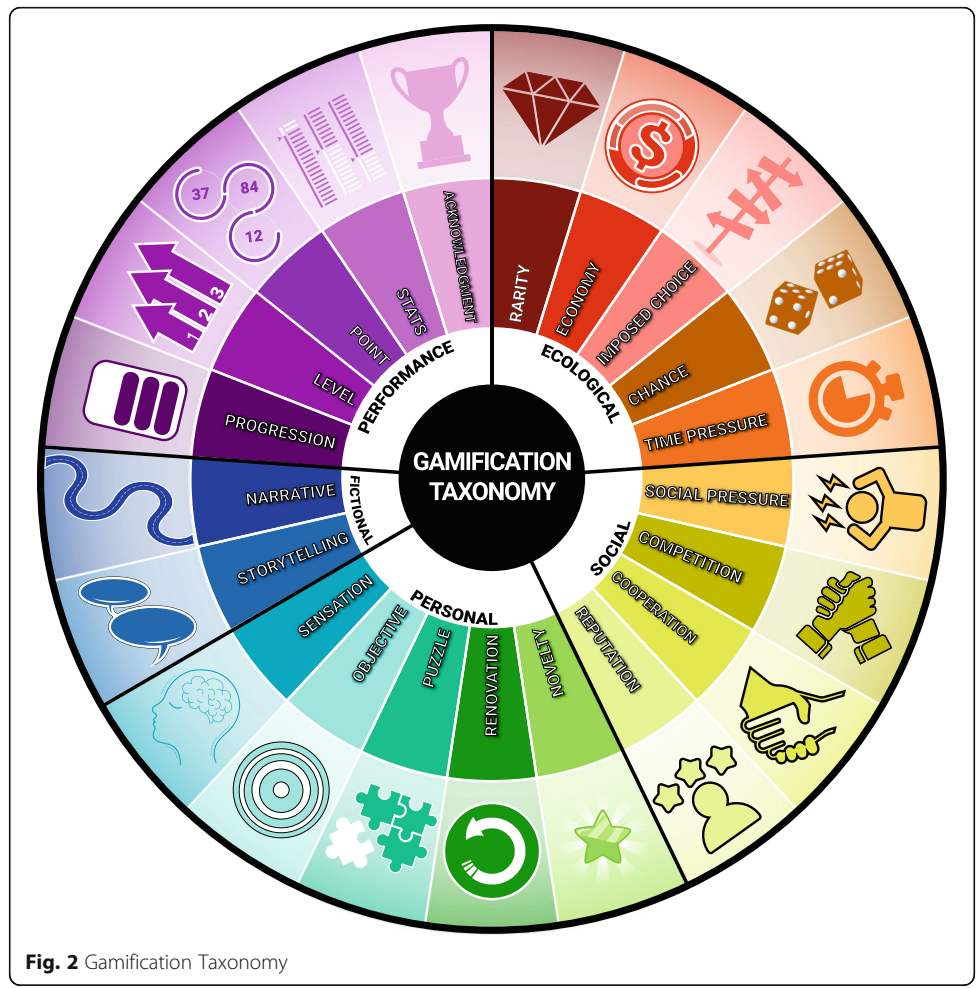
**Fig. 2** Gamification Taxonomy

*Ecological*

This context is related to the environment that the gamification is being implemented. These elements can be represented as properties. The elements in this dimension are Chance, Imposed Choice, Economy, Rarity and Time Pressure. The lack of Ecological elements makes the environment feel dull, as it does not have elements that produce interactions with the user.

- **Chance:** also known as randomness, luck, fortune or probability. This intrinsic concept is related to the random property of a certain event or outcome, e.g. the student may get a random number of points after completing a task; spinning a roulette that may give the user a bonus; user has a probability of getting a special item based on its luck (Dignan 2011).
- **Imposed choice:** also known as choice, judgment, and paths. This extrinsic concept occurs when the player faces an explicit decision that they must make to advance in the environment. An example of this concept is to present the user two different contents and make them choose one or another, blocking their advance if a choice is not to pick.
- **Economy:** also known as transactions, market, exchange. This concept is extrinsically related to any transaction that may occur in the environment.

      Examples are trading points for advantages within the environment and related to the content.
- **Rarity**: also known as limited items, collection, exclusivity. It is related to extrinsically limited resources within the environment which can stimulate the learners through a specific goal.
- **Time Pressure:** also represented as countdown timers or clocks. It is related to time itself used to pressure the learners' actions (extrinsic). In learning environments, this can be represented also as deadlines. It is, alongside Social Pressure, considered one of the most irrelevant elements since it can potentially disengage the learner (Toda et al. 2019b).

*Social*

This dimension is related to the interactions between the learners presented in the environment. The elements in this dimension are Competition, Cooperation, Reputation, Social Pressure. The lack of Social elements can isolate the students, since they will not be able to interact with other students.

- **Competition:** also known as conflict, leader boards, scoreboards, player vs player, etc. It's an intrinsic concept, tied to a challenge where the user faces another user to achieve a common goal, e.g. using scoreboards based on the number of points, badges, levels, etc.
- **Cooperation**: also known as teamwork, co-op, groups, etc. It is also an intrinsic concept (related to a task) where the users must collaborate to achieve a common goal, can be considered the opposite of competition (however, both concepts can be used together). Examples of cooperation are tasks where groups interact with each other and are recognised by these interactions (Shi et al. 2014).
- **Reputation**: also known as classification, status. It is related to titles that the learner may gain and accumulate within the environment (intrinsic). Differing from levels, titles represent more of a social status which does not necessarily reflect on the learners' skills. These titles are usually used within communities to create a hierarchy in the environment.
- **Social Pressure:** Also known as peer pressure or guild missions. This intrinsic concept is related to social interactions that exert pressure on the learner.

*Personal*

This dimension is related to the learner that is using the environment. The elements that are used in this dimension are Sensation, Objective, Puzzle, Novelty, and Renovation. The lack of Personal elements can make the user feel demotivated since the system does not provide meaning for the student.

- **Novelty**: also known as an update, surprise, changes, etc. It is intrinsically related to the updates that occur within the environment, by adding new information, content or even new game elements. It is a good strategy to keep users within the environment to avoid stagnation since longitudinal studies on gamification have shown that a static approach (without updates) may cause disengagement and demotivation (Hanus and Fox 2014).
- **Objectives:** also known as missions, side-quests, milestones, etc. This intrinsic concept is related to goals, it provides the player an end, or a purpose to perform the required tasks. Examples on the use of Objective can be broadened (as getting approved in the course) or more specific (as obtaining a certain score in a task) (Toda et al. 2018a).
- **Puzzle**: also known as challenges, cognitive tasks, actual puzzles, etc. This intrinsic concept is related to the activities that are implemented within the environment, they can be tied or considered as the learning activities since the focus is to provide a cognitive challenge to the learner. This concept is also implicitly present in all educational environments, through quizzes or challenges.
- **Renovation**: also known as boosts, extra life, renewal, etc. This concept is intrinsically related to the property of re-doing a task, event or any of the sorts. It allows the learner a second chance after they fail a task. It is one of the properties that makes games fun (Lee and Hammer 2011).
- **Sensation**: This is either visual or sound stimulation, etc. It is related to the use of learners' senses to improve the experience (intrinsic). This can be done through dynamic and gameful interfaces, Virtual Reality (VR) and/or Augmented Reality (AR).

*Fictional*
It is the mixed dimension that is related to the user (through Narrative) and the environment (through Storytelling), tying their experience with the context. The lack of Fictional elements causes the loss of meaning, of context, that is, the why, within the immersive environment, the user must perform any task, as well as directly influence the quality of the user experience.

- **Narrative:** also known as karma system, implicit decisions, etc. This intrinsic concept is the order of events as they happen in the game, through the user experience. This experience is influenced by implicit choices made by the user. Examples of this are giving a small token of appreciation to the students that opt to interact with other students, subtly and discreetly (Palomino et al. 2019).
- **Storytelling:** can be seen as audio queues, text stories, etc. It is the way the story of the environment is told (as a script). It is told through text, voice, or sensorial resources. It is highly used as a tool to support the narrative within an environment (Palomino et al. 2019).

**Example on the use of the taxonomy**
To demonstrate the analysis and evaluation of these elements, we choose some e-learning environments that are cited and evaluated in the literature (Klock et al. 2017) as Duolingo and MeuTutor. We opted initially for these two for: (a) one is one of the most successful examples of gamification in education and (b) the other due to convenience since we had access to all the functionalities of the system and its design process (convenience sampling).

Duolingo is one of the most famous language apps nowadays and most of its success is due to the gamification that was implemented. According to (Huynh et al. 2016) the main elements of Duolingo are Rewards, Leader-boards, Level-system, and Badges. When analysing these elements, we can observe that other elements from our taxonomy are presented, e.g. Rewards are represented through Lingots, which is a currency obtained when you finish a task. These Lingots can be used in the transaction in the system (**Economy**); the Leader-boards are used to create a **Competition** amongst the user and their friends. The Level-system contains four elements: experience points (**Point**), the content the user chooses when they are learning the language (**Imposed Choice**), the level they are in the language (**Progression**) and the user skills (**Level**). Finally, the Badges are a representation of **Acknowledgement** and can be used in the player profile to increase their **Reputation**. Besides these elements, we can observe all of the Personal dimension since the site appeals visually to the user (**Sensation**), providing them a clear **Objective** (learning languages), achieved through cognitive tasks (**Puzzles**), presenting new content (**Novelty**) and allowing them to redo any task as the user wants (**Renovation**).

By using our taxonomy, we can observe that Duolingo presents a solid Personal (All 5) and Measurement (4 elements) dimensions, some Ecological and Social aspects underneath the system (2 elements each) and no Fiction element.

Following, MeuTutor is an Intelligent Tutoring System used in more than 10 schools in Brazil. The system contains many gamification elements to improve learners' engagement and motivation. According to the description of the system, MeuTutor gamification is based on: learners' being exposed to learning resources and gaining experience points for their interactions (**Point**), which is converted into the learners' **Level**. The content is presented similarly as Duolingo where the learner must choose a topic to continue using the system (**Imposed Choice**). The learners' can see their progress in the course through a completion percentage (**Progression**) and gain badges based on their interactions (**Acknowledgement**), these learners can also see an overall of their performance in a personalised dashboard (**Stats**). The system also presents a leader-board to create a **Competition** amongst its users and collaborative activities, where the learners can create groups to perform a task (**Cooperation**). MeuTutor also presents all the elements in the Personal Dimension, since it has a clear **Objective** (improve learners' knowledge on a certain content), achieved through many learning tasks (**Puzzles**) that can be updated as the teacher desires (**Novelty**). The system also allows the learner to re-do previous tasks (however this **Renovation** does not add to their experience points) and has an interface that is attractive to the final users (**Sensation**).

Through our taxonomy, we can observe that MeuTutor contain a solid Personal and Measurement dimension (All 5 elements of each dimension), it also presents two elements from the Social dimension but does not explore their efficacy, and one element in the Ecological dimension. Also, no Fictional element is present in this environment.

## Discussion

These new dimensions might provide a way to analyse and/or support the design of gamified learning environments. We can also assure that it is aligned with the agenda defined in (Koivisto and Hamari 2019) which states that gamification studies should pay more attention to different types of feedback, as well as exploring and incorporating context and defining universal taxonomies. By creating an initial generalisation of exiting game elements and adapting them to educational contexts, we can infer that instructors and future research may find easier to analyse existing systems and extract the gamification elements within it. Based on the exposed elements, this section will discuss the implications of using each in an educational environment.

### Measurement dimension

This Dimension, as shown in the examples, must always be present so the user may have feedback on their actions. Concerning the game elements in this Dimension, the lack of **Acknowledgement** may lead the user to a state of frustration, since their interactions are not being recognised as something important, whereas providing acknowledgements not properly planned may cause unexpected outcomes (e.g., earning badges based on time to finish a task may lead the students to complete tasks as fast as possible without taking into account whether they are correct). As for **Level**, it is considered a relevant element, especially when tied to **Progression**, according to (Toda et al. 2019a). Lack of levels may lead the learner to think that they did not advance at all in their skills. Following, according to (Toda et al. 2019a), Progression is also considered a highly relevant element to learners, independent of gender. Lack of progression might lead the learner to a feeling of frustration and anxiety (Dignan 2011). Finally, **Stats** is also presented in almost all educational environments, the lack of information makes the learner feel disoriented (Dignan 2011). Although, the literature on this topic still hasn't reached a consensus on the best way to relate or use these elements properly.

### Ecological dimension

Concerning this Dimension, it is related to concepts that act as properties of the environment that can be implemented in a subtle way to engage the users to follow the desired behaviour. They can be supported by the elements in the Measurement/Feedback Dimension, to ensure the behaviour is followed. Although, most of these elements must be designed with care since they can affect the learners' interactions drastically. **Chance** is directly affected by the users' luck; some strategies can be used to mitigate the "bad luck" effect as including an automatic success after a certain number of tries. When the **Economy** is not related to the content, the user may lose focus on what is important (Snow et al. 2015). Although, the users might find attractive when the Economy is tied directly to advantages related to the class, e.g. using their coins to postpone a test (Toda et al. 2016). Lack of **Imposed Choice** within the system might lead the learner to a state where they feel their actions are not meaningful (Dignan 2011), at the same time excessive freedom may allow the students to perform undesired actions. As for **Rarity**, the addition of limited events that rewards exclusive badges or another kind of feedback may engage the learners, but the presence of rare resources and their constraints might demotivate the learners. Lack of rarity may lead the learner to boredom (Dignan 2011) Finally, the absence of **Time Pressure** may lead the learner to a state of boredom since they might not feel challenged or pressured to complete a task (Dignan 2011). An example of using time pressure more healthily is to provide flexible deadlines, where the learner is responsible for the completion of the task. Time pressure is implicit in Massive Online Open Courses (MOOC) systems, which may be one of the reasons why the dropout rates are high (Cristea et al. 2018).

### Social dimension

This Dimension is concerned with the Social aspects of the environment. The elements that connect people and influence their behaviour towards a task. Since this Dimension is concerned with interactions between the learners (instead of interactions with the system) it must be designed carefully. On one hand, **Competition** can create a healthy environment where the learners try to overcome their peers to achieve a certain prize, on the other hand, it has a huge potential to demotivate the learner when their performance is not as expected. An example on how to design a good competition is trying not to tie it to any content-based activity, or by creating groups to mitigate any kind of isolation effect[4] (Papadopoulos et al. 2016; Toda et al. 2018a). **Cooperation** is seen as a positive addition in most educational environments, although its implementation may be complex. The absence of Cooperation may lead to isolation, which may increase the odds of demotivation or disengagement of the learner, whereas using it may lead the students to share knowledge and to work harder in order to avoid jeopardising their peers. One of the major examples of success in Cooperation is Wikipedia. **Reputation** is related to the social status the learners may acquire in the environment, e.g. Best Student in the course. Lack of reputation is similar to the lack of acknowledgment and point, where the learner may feel their actions are not meaningful (Dignan 2011), but also must be designed with care or learners might feel demotivated due to not acquiring a certain status. Finally, **Social Pressure** is usually considered as one of the most irrelevant amongst all elements (Toda et al. 2019a) but can be helpful if properly designed, e.g. persuading a high score learner to encourage a disengaged peer that has a poor performance. Assigning peer-review activities might also imply social pressure.

### Personal dimension

This dimension is directly related to the learner using the environment. It presents elements that are intrinsic to educational environments and the learner might not perceive these elements as gamification. According to our analysis, all educational environments contained the five elements of this dimension, however, some of them (e.g. MeuTutor) did not use these elements in a way that could favour the learner (e.g. the Renovation element is present, but the student does not gain more points by redoing a task). Concerning each element, **Objective** is presented in all educational environments, since the main focus of these applications is to make the student learn or practice a concept, whilst being cautious not to encourage undesired behaviours (e.g., an objective of completing many tasks may lead students to complete numerous of those without properly seeking to correctly complete them). Lack of objectives may misguide or confuse the learner. According to Toda et al. (2019a), it is the most relevant element to use within gamified educational environments. Besides, as repetition or static environments[5] may jeopardise the learning process, **Novelty** may aid in this perspective as well (Mustafa et al. 2019). However, adding Novelty can be complex as it often requires human effort or automatic generation techniques (Shehadeh et al. 2017). As for the **Puzzle**, it is represented through challenges and cognitive tasks, which are common to any educational environment. Lack of Puzzle can make the environment look dull and demotivate the user (e.g. an environment where the learner can only watch videos without any kind of interaction or tasks related to it may lead these learners to boredom.). **Renovation** is also a concept that is present in almost all educational environments since learners can redo a task if they fail, or just want to remember a concept (e.g. rewatching a video or redoing a task). Although, Renovation is not always presented as a gamification element since the user does not benefit from using it (e.g. gaining points by rewatching a video). Lack of Renovation usually make learning environments to feel more difficult which can demotivate the students. In most nonvirtual educational environments, the lack of Renovation is what makes the educational process tedious to learners (Smith-Robbins 2011). Finally, **Sensation** is usually presented as a pleasant interface that is appealing to the user. Although some educational environments are investing in the use of Augmented Reality (AR) and Virtual Reality (VR) with gamification features, it is still a new emergent area. According to Toda et al. (2019a), Sensation is considered a highly relevant element.

---

[4]Where a student feels they are not good enough and stop doing the activities
[5]Environments that do not receive updates

### Fiction dimension

Considering this Dimension, it is not common or considered when designing gamified educational environments (Palomino et al. 2019). This occurs since most gamification frameworks do not make a differentiation between Narrative different layers and Storytelling. The **Narrative** is related to the learner's interaction with the system, affected by their characteristics. If designed correctly, it can help the learner to focus on the content rather than the game elements around it. **Storytelling** is a way to materialise the Narrative, using techniques with the aid of text, audio-visual and another sensorial stimulus, stabilising how the story (or context) is told. The absence of Narrative may hinder the students' engagement and focus on the content to be learned. As for the lack of Storytelling, it might lead to context confusion, causing the student to not see a reason to perform a certain task from the gamification point of view. Storytelling can be used to give a context (e.g. a theme) to the environment, e.g. telling the learner they are fighting a boss that takes damage for each correct task.

### Conclusions and future works

This work presented how we could use an existing taxonomy to analyse and evaluate gamified systems. We improved the description of the game elements as well as provide examples on how to use each to analyse educational systems. We also proposed an initial hierarchy to classify those elements into Five Dimensions, which can provide support to designers and developers of educational environments. Finally, we proposed a link between this hierarchy and aspects such as feedback, user interaction, and motivation.

Through our discussions, we debate some advantages and disadvantages of using each Dimension. Some limitations of the current work are that we did not evaluate the acceptance of the new grouped dimensions with experts, but with five researchers only due to time constraints. Finally, as future work, we intend to explore the learners' perception of this taxonomy to identify the best practices on how to use the elements properly (by using data-driven approaches and/or machine learning algorithms). Through this future exploration, we believe we may find concrete guidelines on how to gamify educational environments and give those guidelines to teachers, instructors, designers and/or developers.

#### Acknowledgements
The authors would like to thank the funding provided by FAPESP (Projects 2016/02765-2; 2018/11180-3; 2018/15917-0; 2018/07688-1), FAPESC (public call FAPESC/CNPq No. 06/2016 support the infrastructure of CTI for young researchers, project T.O. No.: 2017TR1755 - Ambientes Inteligentes Educacionais com Integração de Técnicas de Learning Analytics e de Gamificação) and CNPq. This study was financed in part by the Coordenação de Aperfeiçoamento de Pessoal de Nível Superior - Brasil (CAPES) Finance Code 001.

#### Authors' contributions
The taxonomy was proposed by AMT, ACTK, WO, LR, PTP and AIC. The visual representation was designed by PTP. The study design was proposed by AMT, ACTK, IG and AIC. AIC, SI, IB, LS and IG supervised the activities and wrote parts of and reviewed the paper for final submission. All authors read and approved the final manuscript.

#### Funding
This project was funded by Fundação de Amparo à Pesquisa do Estado de São Paulo (FAPESP), Fundação de Amparo à Pesquisa e Inovação do Estado de Santa Catarina (FAPESC), Conselho Nacional de Desenvolvimento Científico e Tecnológico (CNPq) and Coordenação de Aperfeiçoamento de Pessoal e Nível Superior (CAPES).

#### Availability of data and materials
All the materials and data are available within the paper.

#### Competing interests
The author(s) declare(s) that they have no competing interests.

#### Author details
[1]Universidade de Sao Paulo - Av. Trab. Sao Carlense, 400, zip code, Sao Carlos, SP 13566-590, Brazil. [2]Durham University - Lower Mountjoy South Road Durham, zip code, Durham DH1 3LE, UK. [3]Federal University of Rio Grande do Sul - Av. Bento Gonçalves, 9500, zip code, Porto Alegre, RS 91501-970, Brazil. [4]Santa Catarina State University - Zona Industrial Norte, zip code, Joinville, SC 89219-710, Brazil. [5]Federal Univeristy of Alagoas - Av. Lourival Melo Mota, zip code, Maceió, AL 57072-900, Brazil.

## Publisher's Note

Springer Nature remains neutral with regard to jurisdictional claims in published maps and institutional affiliations.